\documentstyle[12pt,fullpage,psfig]{article}
\begin{document}
\begin{titlepage}
\title{Survival and Extinction in Cyclic and Neutral Three--Species 
Systems}
\author{Margarita Ifti and Birger Bergersen \\
Department of Physics and Astronomy, University of British Columbia \\ 
6224 Agricultural Road, Vancouver, BC, Canada V6T 1Z1}
\maketitle
\normalsize 

\begin{abstract}
We study the $ABC$ model ($A + B \rightarrow 2B$, $B + C \rightarrow 2C$, 
$C + A \rightarrow 2A$), and its counterpart: the three--component 
neutral drift model ($A + B \rightarrow 2A$ or $2B$, $B + C \rightarrow 
2B$ or $2C$, $C + A \rightarrow 2C$ or $2A$.) In the former case, the 
mean field approximation exhibits cyclic behaviour with an amplitude 
determined by the initial condition. When stochastic phenomena are taken 
into account the amplitude of oscillations will drift and eventually one 
and then two of the three species will become extinct. The second model 
remains stationary for all initial conditions in the mean field 
approximation, and drifts when stochastic phenomena are considered. We 
analyzed the distribution of first extinction times of both models by 
simulations of the Master Equation, and from the point of view of the 
Fokker-Planck equation. Survival probability vs. time plots suggest an 
exponential decay. For the neutral model the extinction rate is inversely 
proportional to the system size, while the cyclic model exhibits 
anomalous behaviour for small system sizes. In the large system size 
limit the extinction times for both models will be the same. This result 
is compatible with the smallest eigenvalue obtained from the numerical 
solution of the Fokker-Planck equation. We also studied the behaviour of 
the probability distribution. The exponential decay is found to be robust 
against certain changes, such as the three reactions having different 
rates.

\end{abstract}
\end{titlepage}

\section{Introduction}

Cyclic phenomena are often ignored when studying epidemiological and 
evolutionary processes, but nevertheless they can have important 
consequences, e.g. we contract most infective diseases only once in our 
lifetime, because our immune system has ``memory''. Vaccines are designed 
based on this knowledge, and they work quite efficiently. However, it is 
widely accepted that some viruses, such as the flu virus, mutate in order 
to make themselves unrecognizable by our immune system, and thus be able 
to reinfect us. Immunity to pertussis (whooping cough) is temporary, and 
decreases as the time after the most recent pertussis infection 
increases. The chicken pox (agent VZV---varicella zoster virus) also 
repeats, because immunity wanes with time. Our model is motivated by this 
type of situation.

In epidemiological models populations are often categorized in three 
states: susceptible $(S)$, infected $(I)$ and recovered $(R)$. There 
exists a vast literature about the so-called $SIR$ models, where the 
``loss of immunity'' step is not considered (e.g. the classic texts by 
Bailey~\cite{bib1}, Anderson and May~\cite{bib2}, the review by 
Hethcote~\cite{bib3}). We consider the case when the ``loss of immunity'' 
step is present, otherwise referred to as $SIRS$ models~\cite{bib4, bib5}.

Recently it has been shown that many species of bacteria are able to 
produce toxic substances that are effective against bacteria of the same 
species which do not produce a resistance factor against the 
toxin~\cite{bib6, bib7}. These bacteria exist in colonies of three 
possible types: sensitive $(S)$, killer $(K)$ and resistant $(R)$. The 
killer type produces the toxin, and the resistance factor that protects 
it from its own toxin, at a metabolism cost. The resistant strain only 
produces the resistance factor (at a smaller metabolism cost). The 
sensitive strain produces neither. Clearly, the $S$ colony can be invaded 
by a $K$ colony, the $K$ can be invaded by $R$, and the $R$ can be 
invaded by $S$.

In our first model the population is categorized into three ``species'': 
$A$, $B$, $C$, and the rules are such that when an $A$ meets a $B$, it 
becomes $B$, when $B$ meets $C$, it becomes $C$, when $C$ meets $A$, it 
becomes $A$. The total population size is conserved. Otherwise this could 
be seen as a three-party voter model, when the follower of a certain 
party ``converts'' a follower of another certain party when they meet.

The reaction is similar to the cyclical ``Rock-paper-scissors game'', of 
which a biological example has been found recently~\cite{bib8, bib9}. It 
is played by the males of a lizard species that exist in three versions: 
the blue throat male defends a territory that contains one female, the 
orange throat male defends a territory with many females, and the male 
with a throat with yellow stripes does not defend his own territory, but 
can sneak into the territory of the orange males and mate with their 
females. Hence, a ``blue'' population can be invaded by ``orange'' males, 
while an ``orange'' population is vulnerable to ``yellow'' males, who on 
their turn are at an disadvantage against the ``blue'' males, who defend 
their territory very well. Many spatial models that relate to such 
systems have been built~\cite{bib10, bib11, bib12, bib22}.

The second model is a three--component version of the famous Kimura model 
of neutral genetic drift~\cite{bib15, bib16}. In that case, when 
individual from two species meet, the offspring may be either of the 
first or the second species, with equal probability.

\section{Description of the Model}

Consider a system in which three species $A$, $B$, $C$ are competing in a 
way described by the reaction: $A + B \rightarrow 2B$, $B + C \rightarrow 
2C$, $C + A \rightarrow 2A$. 

The rate equations for this system will be:
\begin{eqnarray}
N \frac {d A}{d t} = AC - AB \nonumber \\
N \frac {d B}{d t} = BA - BC \\
N \frac {d C}{d t} = CB - CA \nonumber
\end{eqnarray}

\noindent with $A + B + C = N =$ const. (we assume the rates are the 
same, in which case a time--rescale will remove them from the equations.) 
These equations can be rewritten as: 

\begin{eqnarray}
N \frac {d}{d t} \ln A = C - B \nonumber \\
N \frac {d}{d t} \ln B = A - C \\
N \frac {d}{d t} \ln C = B - A \nonumber
\end{eqnarray}

\noindent which leads to the second conservation rule: $ A B C = H $ = 
const.

The above model contrasts with the neutral drift model ($A + B 
\rightarrow 2A$ or $2B$, $B + C \rightarrow 2B$ or $2C$, $C + A 
\rightarrow 2C$ or $2A$) for which:

\begin{equation}
\frac {d A}{d t} = \frac {d B}{d t} = \frac {d C}{d t} = 0
\end{equation}

Assuming that the system is subject to stochastic noise due to Poisson 
birth and death processes (intrinsic noise) we get the master equation 
for the cyclic model:

\[
\frac{\partial P(A,B,C,t)}{\partial t} = \frac {1}{N} [(A-1)(C+1) 
P(A-1,B,C+1,t) - AC P(A,B,C,t) +\]\[ + (A+1) (B-1) P(A+1,B-1,C,t) - 
ABP(A,B,C,t) +\]\begin{equation}
+ (B+1)(C-1) P(A,B+1,C-1,t) - BCP(A,B,C,t)]
\end{equation}

For the neutral drift case the master equation is:

\[
\frac{\partial P(A,B,C,t)}{\partial t} = \frac {1}{N} [\frac {1}{2} 
(A-1)(C+1) P(A-1,B,C+1,t) +\]\[+ \frac {1}{2} (A+1)(C-1) P(A+1,B,C-1,t) - 
AC P(A,B,C,t) +\]\[ + \frac {1}{2} (A+1) (B-1) P(A+1,B-1,C,t) + \frac 
{1}{2} (A-1)(B+1) P(A-1, B+1,C,t) -\]\[ - ABP(A,B,C,t) + \frac {1}{2} 
(B+1)(C-1) P(A,B+1,C-1,t) + \]\begin{equation} + \frac {1}{2} (B-1)(C+1) 
P(A,B-1,C+1,t) - BCP(A,B,C,t)]
\end{equation}

Now we introduce the ``shift'' operators, defined by

\begin{equation}
\epsilon_{A} f(A,B,C) = f(A+1,B,C)
\end{equation}

and

\begin{equation}
\epsilon^{-1}_{A} f(A,B,C) = f(A-1,B,C)
\end{equation}

\noindent and likewise for B and C operators. The master equation for the 
cyclic model now reads:

\begin{equation}
\frac {\partial P(A,B,C,t)}{\partial t} = \frac {1}{N}[(\epsilon_C 
\epsilon ^{-1}_A -1) AC + (\epsilon_A \epsilon^{-1}_B -1)AB +(\epsilon_B 
\epsilon^{-1}_C -1)BC] P(A,B,C,t)
\end{equation}

\noindent and for the neutral drift one:

\[
\frac {\partial P(A,B,C,t)}{\partial t} = \frac {1}{N}[(\frac 
{1}{2} (\epsilon_C \epsilon ^{-1}_A + \epsilon_A \epsilon^{-1}_C) -1) AC + 
(\frac {1}{2} (\epsilon_A \epsilon^{-1}_B+\epsilon_B \epsilon^{-1}_A)-1)AB 
+ \]\begin{equation} +(\frac {1}{2} (\epsilon_B \epsilon^{-1}_C + 
\epsilon_C \epsilon^{-1}_B) -1)BC] P(A,B,C,t)
\end{equation}

Next we transform to the intensive quantities

\begin{equation}
x = \frac {A}{N}, y = \frac {B}{N}, z = \frac {C}{N}
\end{equation}

We use the system size expansion of Horsthemke and Brenig~\cite{bib17, 
bib18}. However, the notation and style is closer to Van 
Kampen~\cite{bib19, bib20}.

The shift operators become

\begin{eqnarray}
\epsilon_{A(B,C)} = \sum_j \frac {1}{j!} N^{-j} \frac 
{\partial^j}{\partial x^j (y^j, z^j)} \\
\epsilon^{-1}_{A(B,C)} = \sum_j \frac {-1^j}{j!} N^{-j} \frac 
{\partial^j}{\partial x^j (y^j, z^j)} \nonumber
\end{eqnarray}

Further we use the rules of transformation of random variables to define:

\begin{equation}
W(x,y,z,t) = P(A,B,C,t) \cdot N^3
\end{equation}

Before we go ahead with the expansion, a few comments are necessary. From 
the rate equations, it is clear that our system does not have one single 
steady state or limit cycle. Instead, the mean-field solution is 
completely dependent on the initial conditions, and we do not have a 
macroscopic solution to expand about, but rather an infinity of neutrally 
stable cycles (for the competition case) or points (for the neutral 
model.) The above expansion, otherwise known as Kramers-Moyal expansion, 
is discussed by van Kampen~\cite{bib19, bib20}. It is a risky expansion, 
and it does not work in most cases, mainly because higher derivatives are 
not small themselves. Also, with time, large fluctuations may occur, and 
since our system presents itself as a fluctuations--driven one, it 
becomes even more important to watch out for this sort of complications. 
In the case of neutral drift, the second order term is the leading one, 
so we would expect this Ansatz to work for large system sizes. In the 
cyclic competition model, the first order term does not drive the system 
out of the (neutrally) stable trajectories, and again it would be the 
second order term to essentially determine the fate of the system. So, we 
set out to investigate the applicability of the Kramers-Moyal expansion 
to the cyclic competition and neutral drift three-species systems.

We transform the master equation into an equation for $W(x,y,z,t)$ 
by substituting the expressions for the shift operators in it, and 
further by grouping together the terms of the same order in $N$. The 
terms of order $N^1$ cancel, and the first terms in the expansion are of 
order $N^0$; keeping only those gives us the first order equation for the 
cyclic case:

\begin{equation}
\frac {\partial W}{\partial t} = [(\frac {\partial}{\partial x} - \frac 
{\partial} {\partial z}) x z + (\frac {\partial}{\partial y} - 
\frac {\partial}{\partial x}) y x + (\frac {\partial}{\partial z} 
- \frac {\partial}{\partial y}) z y] W  
\end{equation}

For the neutral drift case the first-order term is identically zero.

The second order term is obtained by considering the terms of order 
$N^{-1}$ in the expansion of the master equation:

\[\frac {1}{2N} [(\frac {\partial^2}{\partial x^2} -2 \frac {\partial^2} 
{\partial x \partial z} + \frac {\partial^2}{\partial z^2}) x z 
+ [(\frac {\partial^2}{\partial x^2} -2 \frac {\partial^2}{\partial x 
\partial y} + \frac {\partial^2}{\partial y^2}) x y + \]
\begin{equation}
+[(\frac {\partial^2}{\partial y^2} -2 \frac {\partial^2}{\partial y 
\partial z} + \frac {\partial^2}{\partial z^2}) y z]W
\end{equation}

\noindent and is identical for both cyclic and neutral drift cases.

\section{The First Order Term}

The concentrations $x$, $y$, $z$ of a three--component system are 
commonly represented by the distances of a point inside an equilateral 
triangle of unit height from its sides (Fig.~1):

For the neutral drift case the first order equation tells us that in the 
mean--field approximation the system will remain in the initial state 
forever.

After some algebra, the first order equation [13] for the cyclic system 
becomes:

\begin{equation}
\frac {\partial W}{\partial t} = \frac {\partial}{\partial x} (xz-xy)W + 
\frac {\partial}{\partial y} (xy-yz)W + \frac {\partial}{\partial z} 
(yz-xz)W
\end{equation}

At the centre of the triangle ($x=y=z=1/3$) all three expressions above 
are zero, so in the first order approximation the system will stay at 
that state forever. It can be verified that the product $H=x y z$ is a 
solution of the first-order equation, as is any function of that product. 
The lines $x y z = $ const. will then represent possible trajectories of 
the system, which are closed, since there is no term to drive the system 
out of those trajectories. In this (mean-field) approximation, the cyclic 
model exhibits global stability with constant system size 
$N$~\cite{bib28, bib29}. The populations of each species oscillate out of 
phase with one-another, and the amplitude remains constant~\cite{bib21}. 
These are the same trajectories that are obtained when one solves the 
rate equations [1]. Fig.~2  shows some of those closed trajectories.

\section{The Second Order Term}

The second order term [14], which represents the ``diffusion term'' must 
be added to the first order equation to give the Fokker-Planck equation 
for our system. This term is identical for both the cyclic (competing 
species) and neutral genetic drift cases. As we will see, this makes the 
long (evolutionary) time fate of both systems  essentially the same, at 
least for reasonably large system sizes.

The trajectories of a given realisation of the system can easily be 
obtained by simulation: they initially spiral out of the centre, with the 
populations of each species oscillating out of phase with one--another, 
so that the total size of the population is conserved. The amplitude 
drifts until one of the populations becomes zero (species goes extinct). 
In other words, the slow second order term makes the system cross from a 
closed (mean-field) trajectory to a neighbouring one. This would suggest 
an adiabatic approximation. In the neutral drift model, the fluctuations 
drive the system from one point--solution of the mean-field equations to 
a neighbouring one.

The computer simulations of both systems start with an equal number of 
$A$, $B$, $C$, (start at the centre) and generate times for the next 
possible reaction event with exponential distribution as $- N \cdot \ln 
(rn)/AB$, (for $A+B \rightarrow 2B$ reaction, and similarly for the other 
two reactions,) where $rn$ is a random variable with uniform distribution 
in $[0,1]$ (this means the events are really independent)~\cite{gibson}. 
The reaction which occurs first is then picked and the system is updated. 
The process is repeated until one of the species, and then another one, 
go extinct. Fig.~3 shows the variation with time of the number of $A$ for 
a realization of the system in the cyclic competition scenario, and 
Fig.~4 shows the evolution of the number of $A$ population in the neutral 
drift case (in both realizations the total population size is $N=600$.) 
From the time series for the population numbers it looks as if the 
neutral drift model is some sort of ``adiabatic'' approximation of the 
cyclic competition model. If in the neutral drift model the population 
number is random walking, in the cyclic competition one the amplitude of 
oscillations is random walking! In other words, if we average the cyclic 
competition model over the cyclic orbits, we get essentially the same 
behaviour as the neutral drift model. Then the fluctuations drive both 
models from a neutrally stable cycle (point) to the neighbouring ones, and 
these fluctuations remain quite small by virtue of the vicinity of the 
neighbouring cycle (point.) It is quite funny how we all tend to go around 
in circles in life, even though the point we end up to is still the same!

We investigated the probability distribution of first extinction times. 
For that we ran fifty thousands copies of the system for each  
(different) size of the system, and the number of survivors was plotted 
vs. time. In the semilog axes we get a straight line, except for the few 
``rare events''. The slope of this straight line is -3.01 for system size 
3000, -2.98 for system size 6000, -2.99 for system size 9000. Fig.~5 
shows the plots for these three different system sizes: 3000, 6000, and 
9000.

The time dependence of survival probability scales approximately with 
$1/N$, but there is still a very weak dependence left for small $N$. This 
can be justified by looking at the Fokker-Planck equation for this 
system, which contains both a term of order $N^0$ and $N^{-1}$.

To further investigate the behaviour of the system, we looked at the 
cummulative, conditional on being alive (which is equivalent to a 
normalization condition,) probability distribution for $H=xyz$. If 
$P(H<h|$alive$)$ increases linearly with $H$ for small $H$, the 
extinction rate will be non-zero, and we will get exponential decay. This 
corresponds to a constant distribution for $P(H|$alive$)$. Fig.~6 shows 
these plots for $t=1.25$, $t=1.5$, $t=1.75$, and $t=2.0$ in linear axes.

For the neutral drift case the data from the simulations give us a slope 
of -2.992 for system size 6000, -2.991 for system size 3000. The data 
collapse when time is rescaled with $1/N$, and that is in agreement with 
the fact that there is only a term of the order $1/N$ present in the 
Fokker-Planck equation. In Fig.~7 we show the number of survivors vs. 
time plots for cyclic (competition) and drift cases together, for system 
size $N=6000$. They clearly agree.

\section{Probability Distribution at Long Times}

In order to get an expression for $W$ for long times, we looked at 
our ``experimental'' data. It is possible to find a complete set of 
eigenfunctions of the Laplace equation, which vanish at the boundary. 
Having this complete set of eigenfunctions for the equilateral 
triangle~\cite{bib23, bib24}, the probability density can be written:

\begin{equation}
W = \sum_{m,n} c_{m,n} {\phi}_{m,n}
\end{equation}

\noindent where ${\phi}_{m,n}$ are the abovementioned symmetric 
eigenfunctions. The non-normalized eigenfunctions are then obtained from 
the general expression:

\begin{equation}
\phi_{(m,n)} (x,y) = \sum_{(m,n)} \pm \exp [{\frac { 2 \pi i}{3}} (nx + my)]
\end{equation}

\noindent where we are using the relation $x+y+z=1$ and summation over 
the index pair $(m,n)$ means summation for $(-n,m-n)$, $(-n,-m)$, 
$(n-m,-m)$, $(n-m,n)$, $(m,n)$, $(m,m-n)$~\cite{bib24}.

We get a symmetric eigenfunction when $m+n$ is a multiple of 3, $m$ is 
also a multiple of 3, but $m \neq 2n$, and $n \neq 2m$. \footnote{In 
[24], of all the symmetric eigenfunctions, Pinsky only considers the 
non-degenerate ones. There are also two-fold degenerate symmetric 
eigenfunctions, which we verified to be orthogonal to the original 
eigenfunctions.}

In order to calculate the expansion coefficients, we took snapshots of 
the system at given times, and then used the relation:

\begin{equation}
c_{(m,n)} (t) \propto \frac {1}{p} \sum_{l=1}^p { \phi_{(m,n)} (x_l(t), 
y_l(t))}
\end{equation}

\noindent where $p$ is the number of experimental points. The 
time-evolution of the expansion coefficients is given in Fig.~8. The 
doubly--degenerate functions die out quite soon, while the Lam\'e 
symmetric functions persist.

With the coefficients obtained this way we can then construct the 
probability density at different times. Some snapshots at the $W(t)$ are 
shown in Fig.~9.

At $t=0$ the $W(0)$ is a $\delta$-function. As time passes, for $t=0.1$ 
we see the $W$ starts to spread, and it spreads even more for $t=0.2$, 
becoming a ``cake'' for very long times ($t=1.5$,) for both the cyclic 
system and the neutral drift one. The ``cake'' obtained this way shows 
some Gibbs oscillations, which are due to the inclusion of only a few 
terms in the expansion, and the presence of the absorbing boundary. For a 
uniform probability distribution the expansion coefficients are 
proportional to $1/m$, where $m$ is the first index in the pair $(m,n)$. 
It can be seen that our ``experimental'' expansion coefficients approach 
those values.

The second--order Fokker--Planck equation for the drift case accepts 
solutions of the form:

\begin{equation}
W(x,y,z,t) = e^{-\lambda t} W(x,y,z)
\end{equation}

\noindent with $\lambda$ the smallest eigenvalue of the spatial equation. 
We solved the eigenvalue problem by using the Galerkin method, with the 
help of Maple, and obtained a value of $\lambda = 3.01$ for the smallest 
eigenvalue, when we keep the first six eigenfunctions in the expansion. 
This is in very good agreement with our ``experimental'' data.

\section{Case When Rates Are Not the Same}

Now consider the cyclic (competition) case when the rates of the three 
equations are not the same, i. e. the rate equations read like:

\begin{eqnarray}
N \frac {d A}{d t} = c_{13} AC - c_{12} AB \nonumber \\
N \frac {d B}{d t} = c_{21} AB - c_{23} BC \\
N \frac {d C}{d t} = c_{32} BC - c_{31} AC \nonumber
\end{eqnarray}

We can always rescale the $A, B, C$ so that the matrix $c_{ij}$ be 
symmetric, and the equations read:

\begin{eqnarray}
N \frac {d A}{d t} = A (\beta C - \gamma B) \nonumber \\
N \frac {d B}{d t} = B (\gamma A - \alpha C) \\
N \frac {d C}{d t} = C (\alpha B - \beta A) \nonumber
\end{eqnarray}

Again, $A+B+C=N=$const. By manipulating the rate equations, we can find 
the second integral of motion to be $A^{\alpha} B^{\beta} C^{\gamma} 
=H=$const. The new centre will then be not at the point (1/3, 1/3, 1/3), 
but at $A = \alpha, B= \beta, C= \gamma$. The lines along which the 
quantity $A^{\alpha} B^{\beta} C^{\gamma}$ is constant are shown in 
Fig.~10.

We need to check whether the exponential decay behaviour is universal 
when the symmetry is broken this way. For this we ran fifty thousand 
simulations for different combinations of $\alpha$, $\beta$, and 
$\gamma$, for system size $N=3000$, starting from the new centre ($A=N 
\cdot \alpha$, $B=N \cdot \beta$, $C=N \cdot \gamma$.) The number of 
survivors vs. time plots were again obtained, and those plots show that 
the exponential decay behaviour is robust. The time scale is now 
dependent on the initial distance from the boundary, with the equal rates 
case having the largest distance, and so the largest time scale (and the 
smallest slope.) Some of those plots are shown in Fig.~11. In that 
figure, the times for the equal rates case (for which the sum of rates is 
3) are multiplied by three, to make them comparable to the times for the 
unequal rates case, for which the sum of rates is 1.

\section{Conclusions}

We have considered an $ABC$ model in both the cyclic competition and 
neutral genetic drift versions, and studied the long-term behaviour of 
such a model. The number of the $A, B, C$ species in the cyclic 
competition case oscillates with a drifting with time amplitude, until 
one of the species (and then the next one in the cycle) goes extinct. In 
the neutral drift case the number of the $A, B, C$ species drifts, until 
one of them becomes zero. In both scenarios the number of survivors vs. 
time plots show an exponential decay, with the same exponent. The result 
is verified by writing and solving the Fokker--Planck equation for the 
second model. Finally, its robustness is checked against variations in 
the rate of the three different reactions in the system. It is very 
interesting to note that there is no difference in the time scale for the 
ensemble of species in cyclic competition and the case of neutral genetic 
drift.

There is growing concern about the effects of habitat fragmentation in 
the survival of the species~\cite{bib13}. If the population of a certain 
species goes extinct in one patch (e.g. a herd, school, swarm) while it 
still survives in other patches, then what is known as ``rescue effect'' 
can prevent global extinction~\cite{bib14, bib25, bib26}. Otherwise, the 
species is doomed to go extinct altogether. The models with individuals 
of ``species'' $A$, $B$, and $C$ distributed in a lattice have been 
studied by Sz\'abo et al.~\cite{bib10, bib11, bib22}. If we try to model 
habitat fragmentation as an ensemble of patches, each of those patches 
can be considered as one copy of our non-spatial system. However, with 
continuous migration in and out of the patch, the non-spatial picture 
described in this paper would be seriously perturbed. One aspect of this 
immigration and emigration has been recently discussed by Togashi and 
Kaneko~\cite{bib30}. In a first approximation, the broad range of 
extinction times could relate to the persistence of species that exist in 
different ``versions'' such as certain lizards, or even some kinds of 
bacteria. Applied to its epidemiological scenario, it would relate to the 
endemicity of certain diseases, namely those with mutating pathogen, and 
then the outbreaks of epidemics in certain patches (areas). The many 
patches will then form a network~\cite{bib27}. We have work in progress 
which places our $ABC$ ensemble in one of these small worlds with a 
scale-free topology.

\section{Acknowledgements}

We thank Nicolaas G. van Kampen for constructive criticism, and Michael 
D\"obeli for helpful and interesting discussions and suggestions.

\begin{figure}
\centerline{\psfig{file=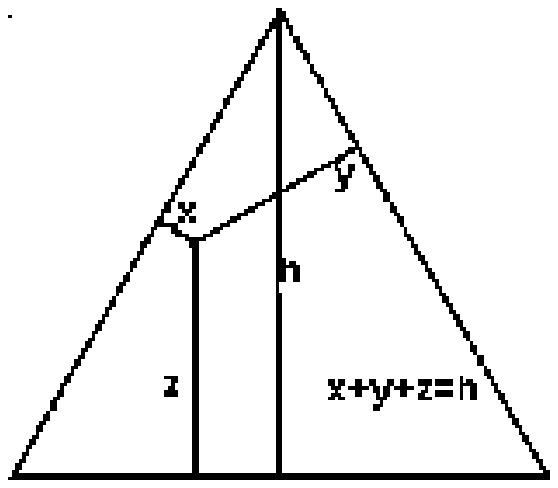}}
\caption{In the equilateral triangle the sum of distances of any point 
from the sides of the triangle is equal to the height of the triangle 
(unity.)}
\end{figure}

\begin{figure}
\centerline{\psfig{file=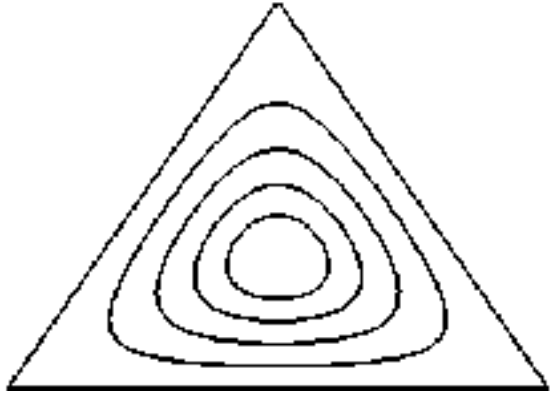}}
 \caption{Some of the closed trajectories of the cyclic competition 
system when only the first order term is considered.}
\end{figure}

\begin{figure}
\centerline{\psfig{file=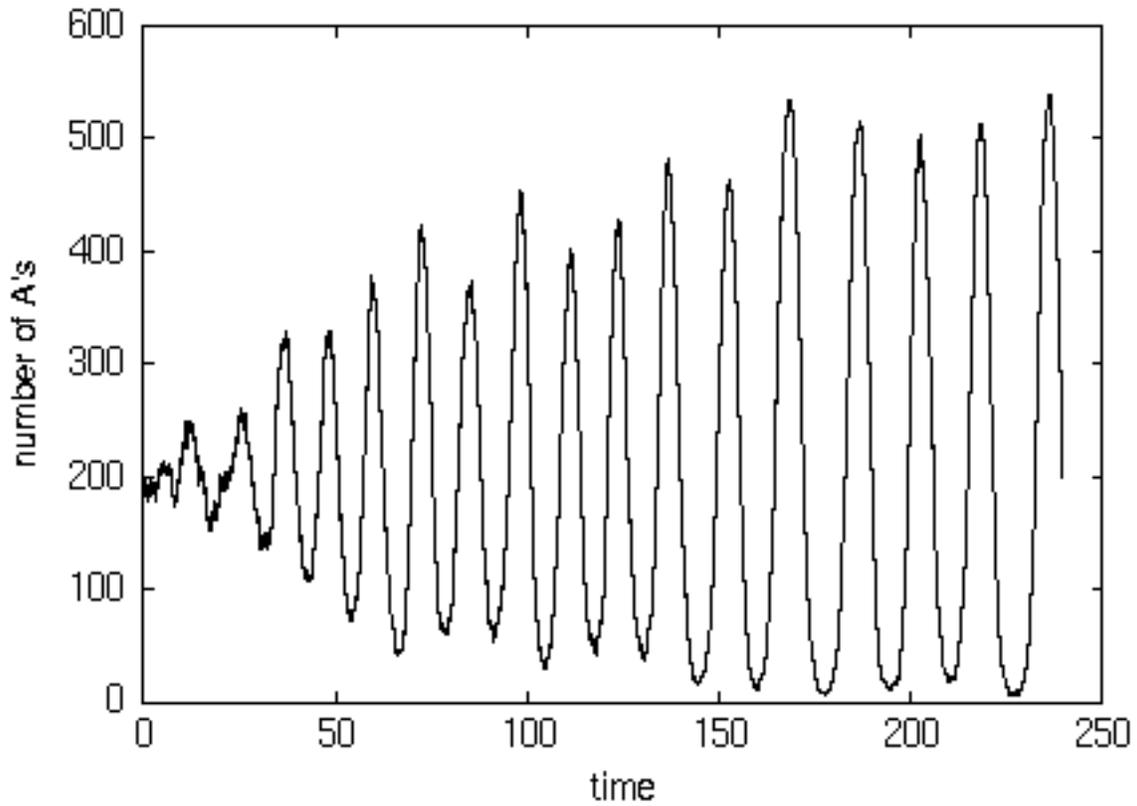}}
\caption{Variation with time of the total number of the $A$ species for a 
realisation of the cyclic competition system ($N=600$).}
\end{figure}

\begin{figure}
\centerline{\psfig{file=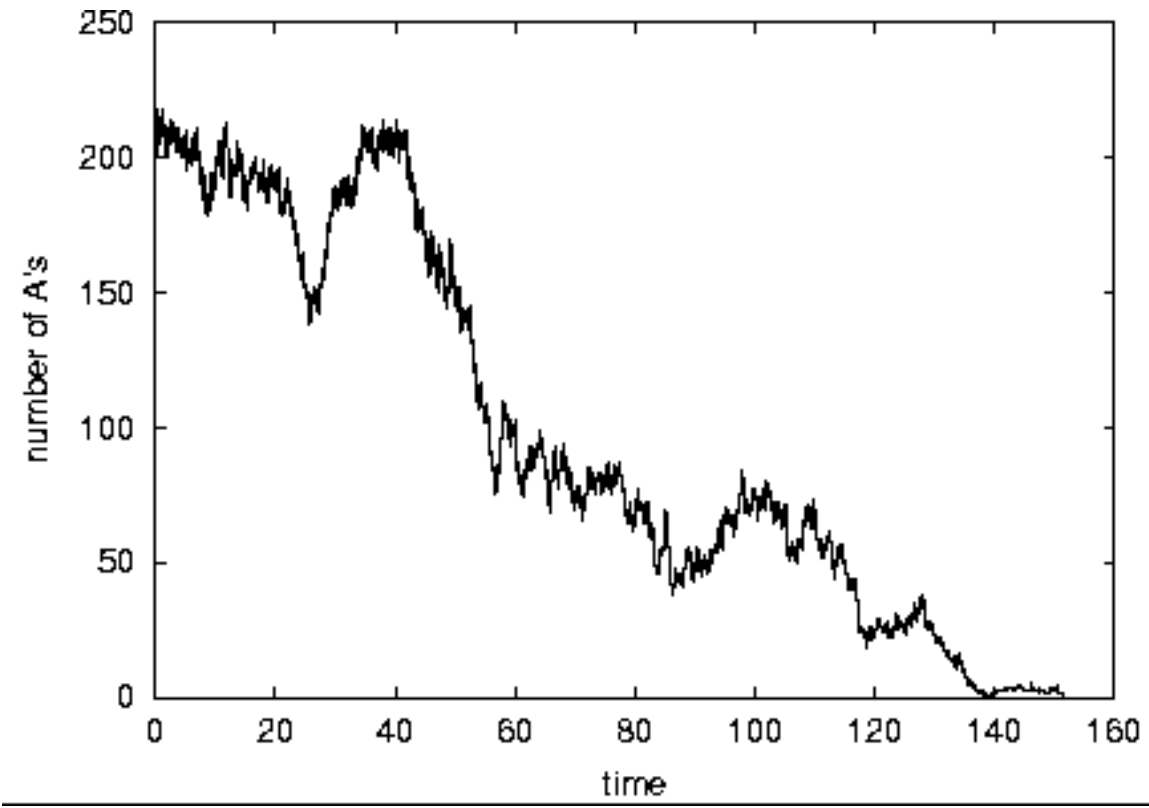}}
\caption{Variation with time of the total number of the $A$ species for a 
realisation of the neutral drift system ($N=600$).}
\end{figure}

\begin{figure}
\centerline{\psfig{file=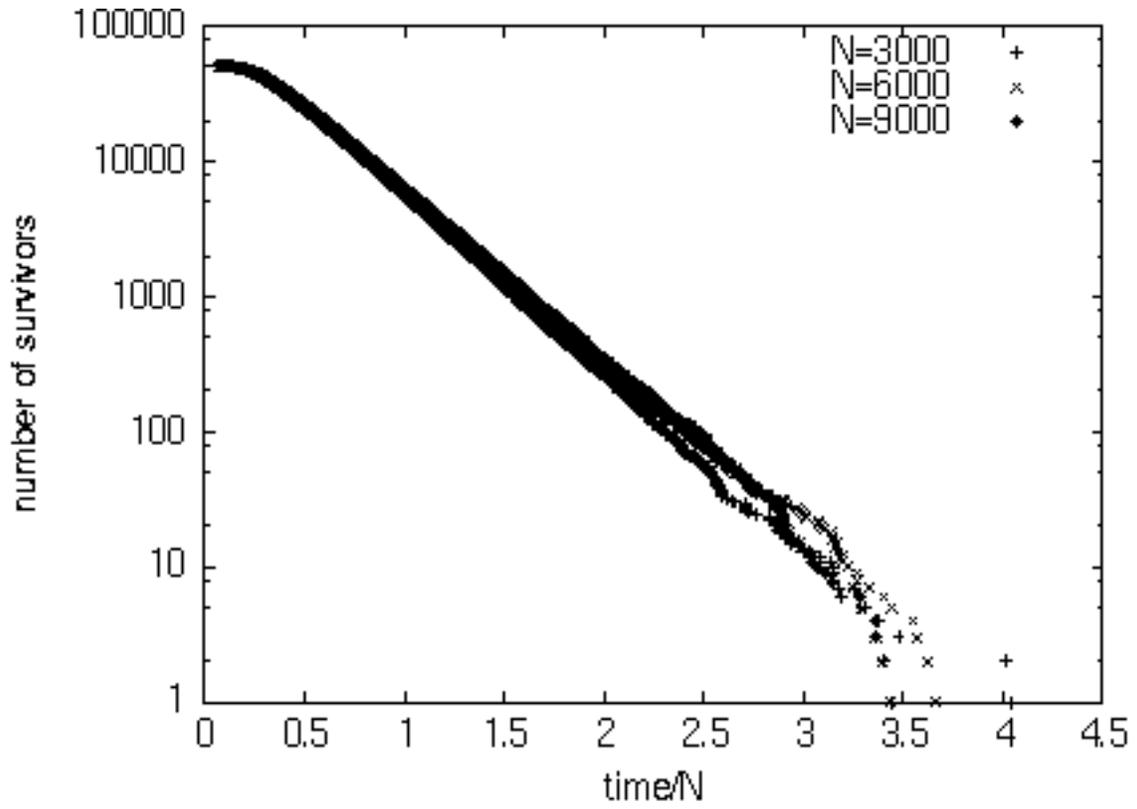}}
\caption{Number of survivors vs. time plots for three different system 
sizes, same rate cyclic competition case.}
\end{figure}

\begin{figure}
\centerline{\psfig{file=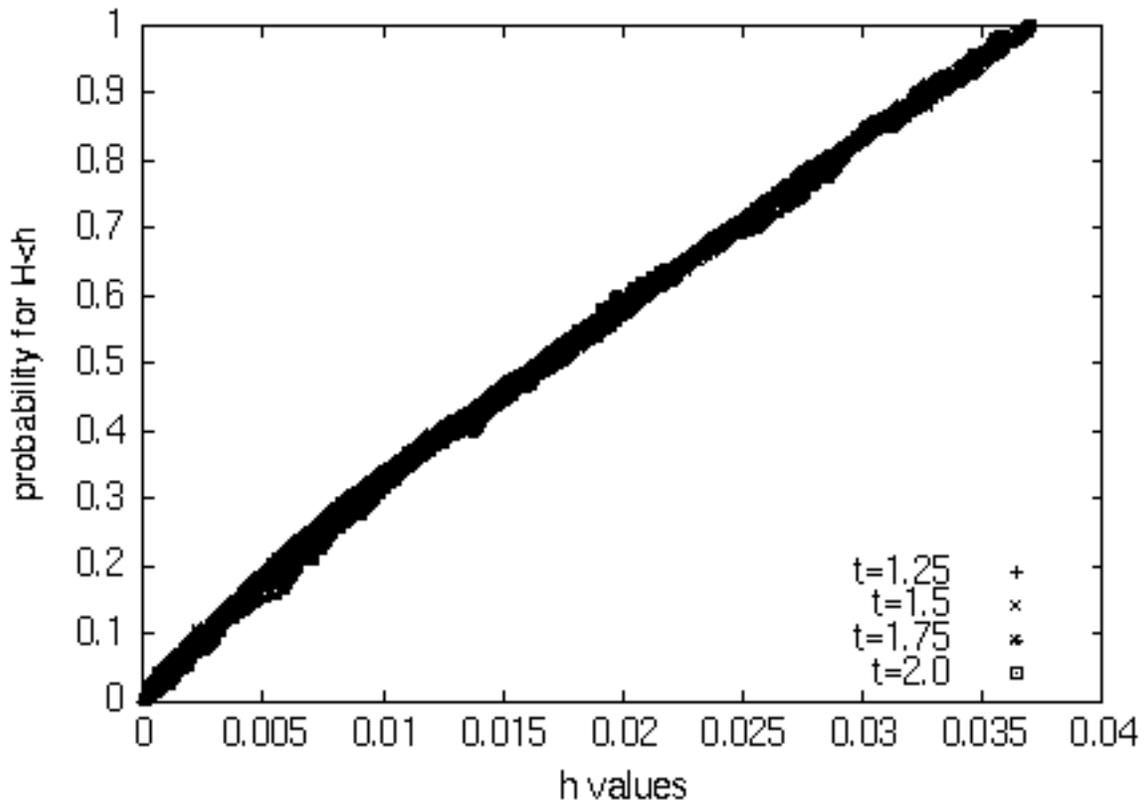}}
\caption{Normalized cummulative probability distribution for the $H=xyz$ 
quantity for the cyclic competition system at $t=1.25$, $t=1.5$, 
$t=1.75$, $t=2.0$.}
\end{figure}

\begin{figure}
\centerline{\psfig{file=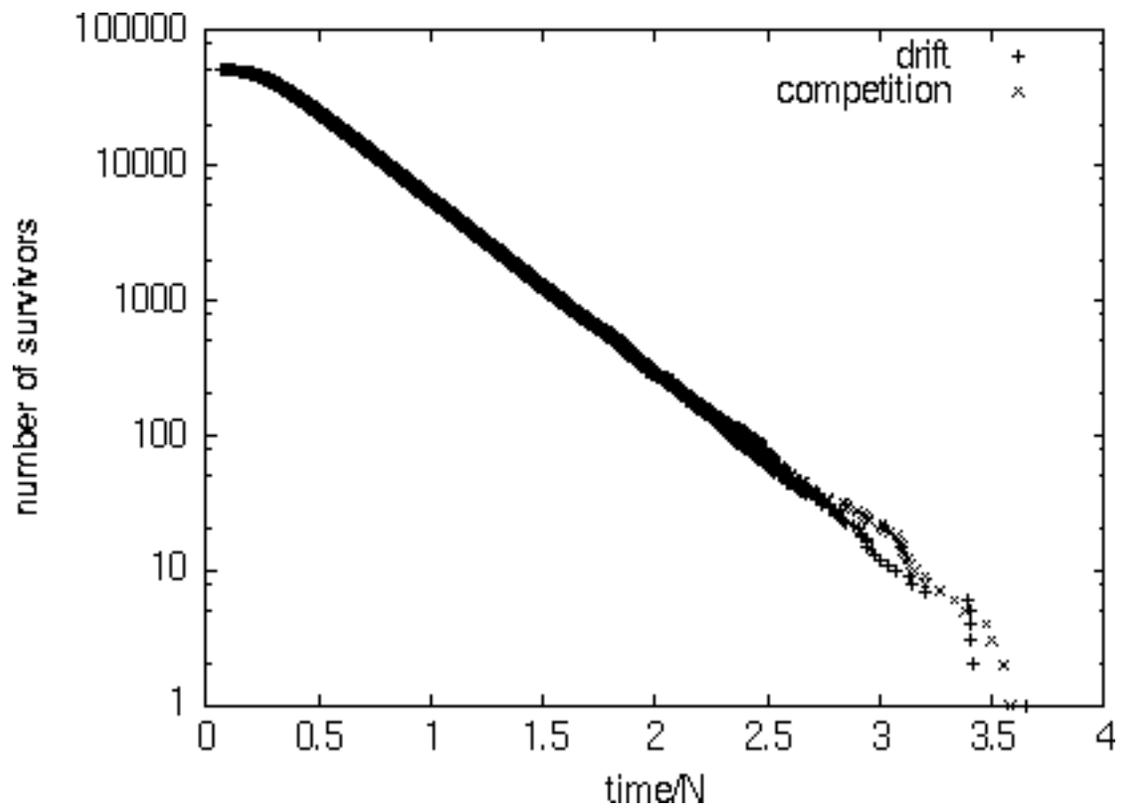}}
\caption{Number of survivors vs. time plots for the cyclic and drift case 
for system size $N=6000$.}
\end{figure}

\begin{figure}
\centerline{\psfig{file=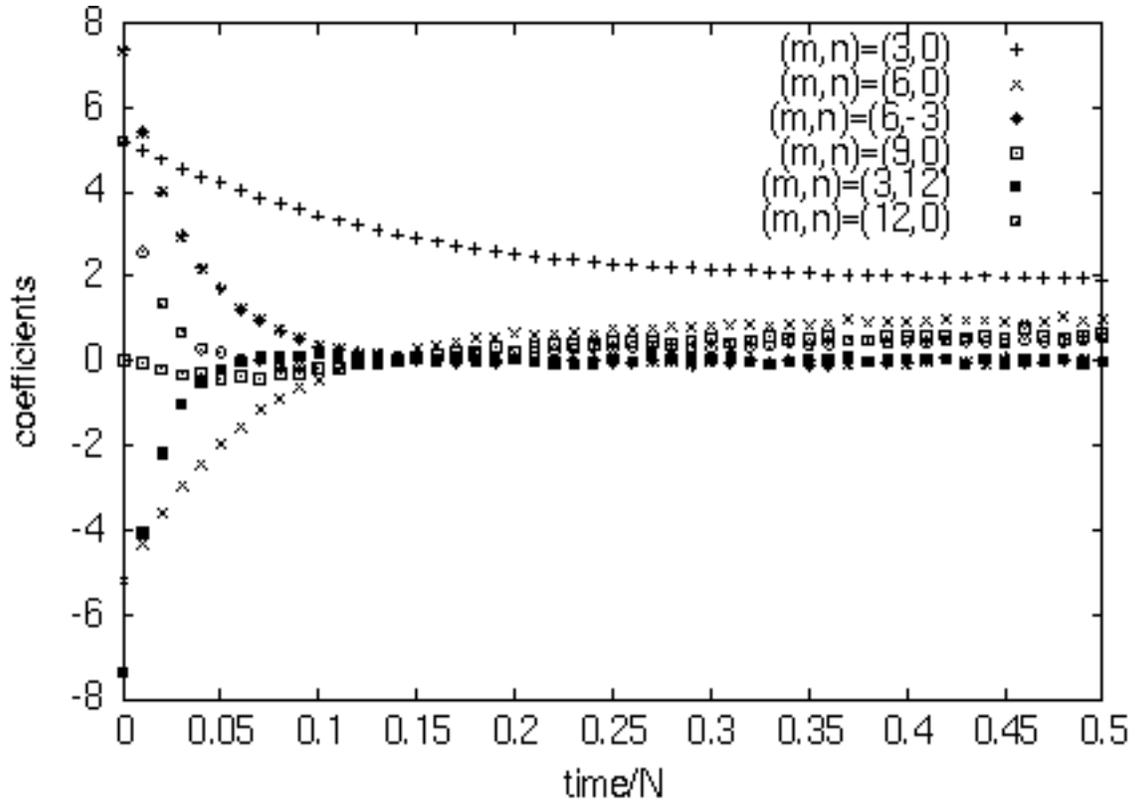}}
\caption{The variation with time of the coefficients for the first six 
functions in the expansion of $W(t)$: $(m,n)=(3,0)$, $(m,n)=(6,0)$, 
$(m,n)=(6,-3)$, $(m,n)=(9,0)$, $(m,n)=(3,12)$, $(m,n)=(12,0)$.}
\end{figure}

\begin{figure}
\centerline{\psfig{file=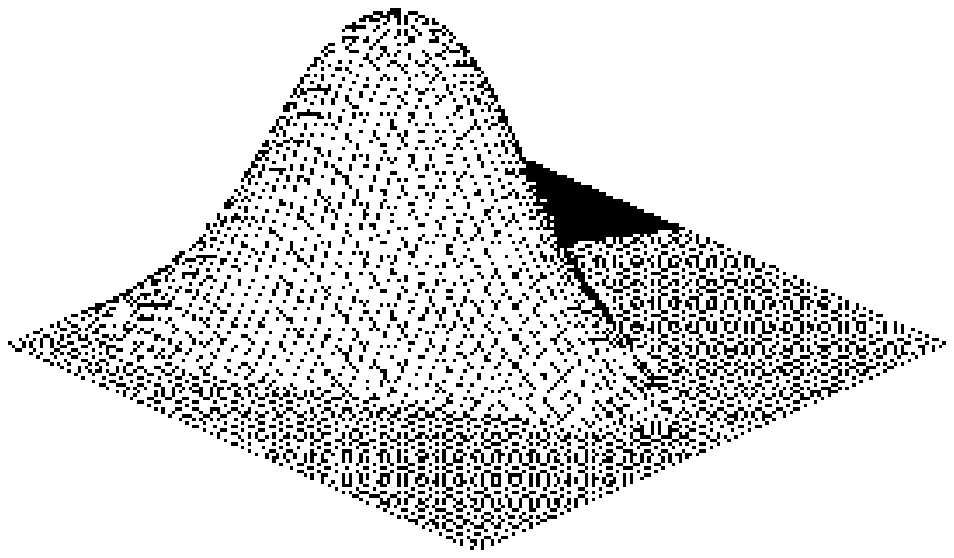}}
\centerline{\psfig{file=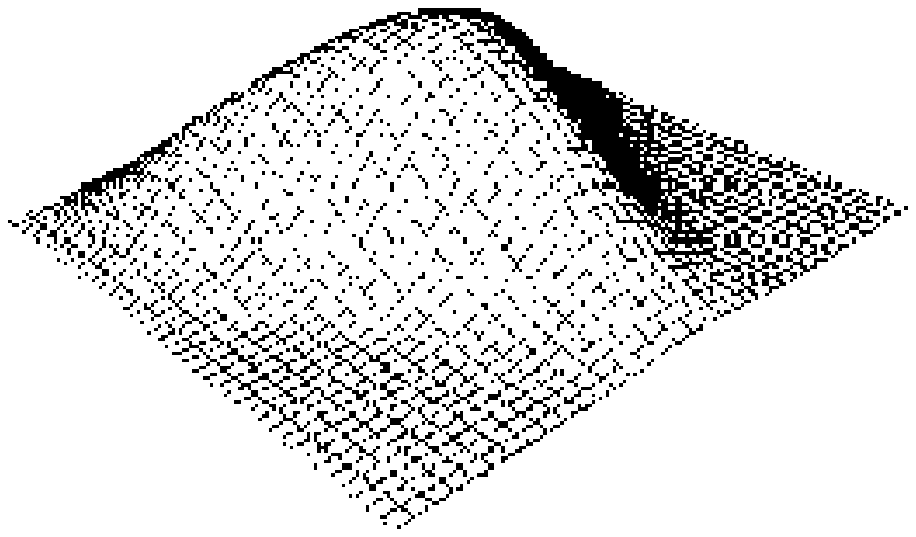}}
\centerline{\psfig{file=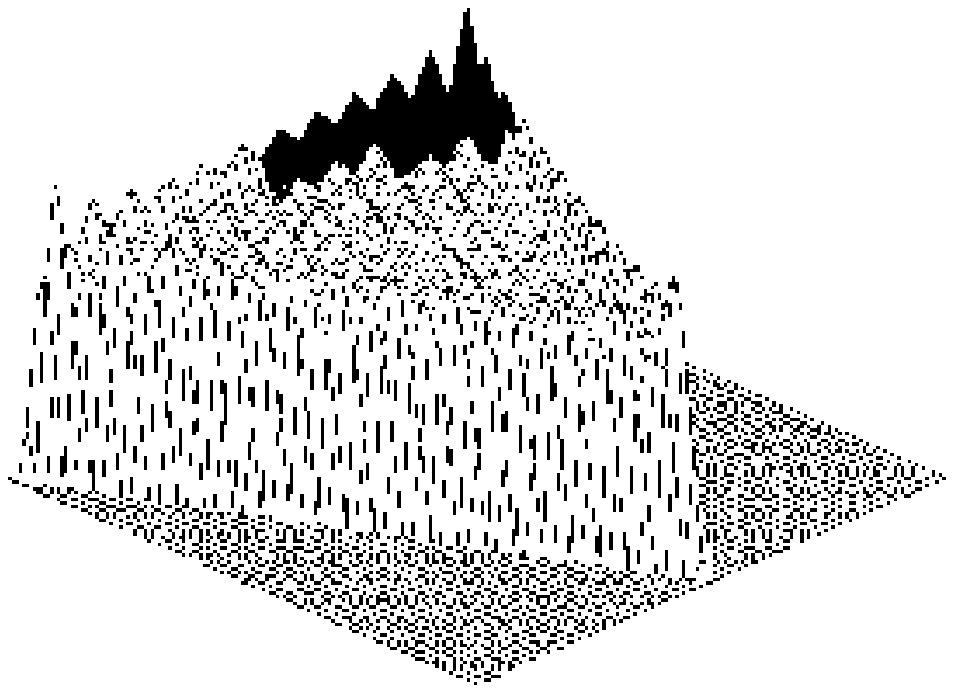}}
\caption{Snapshots at the probability density function at $t=0.1$, 
$t=0.2$, and $t=1.5$.}
\end{figure}

\begin{figure}
\centerline{\psfig{file=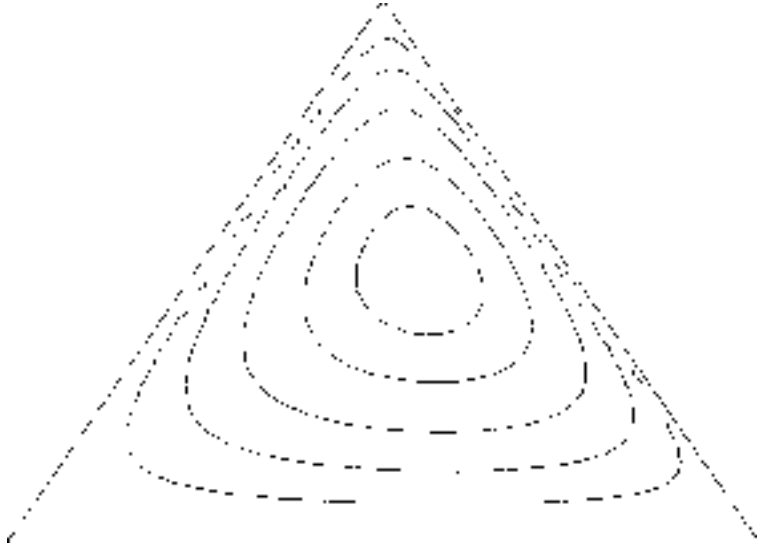}}
\caption{The $H=$const. lines for the case when three reactions happen at 
different rates. Here $\alpha=$0.2, $\beta=$0.3, 
$\gamma=$1$-\alpha-\beta=$0.5.}
\end{figure}

\begin{figure}
\centerline{\psfig{file=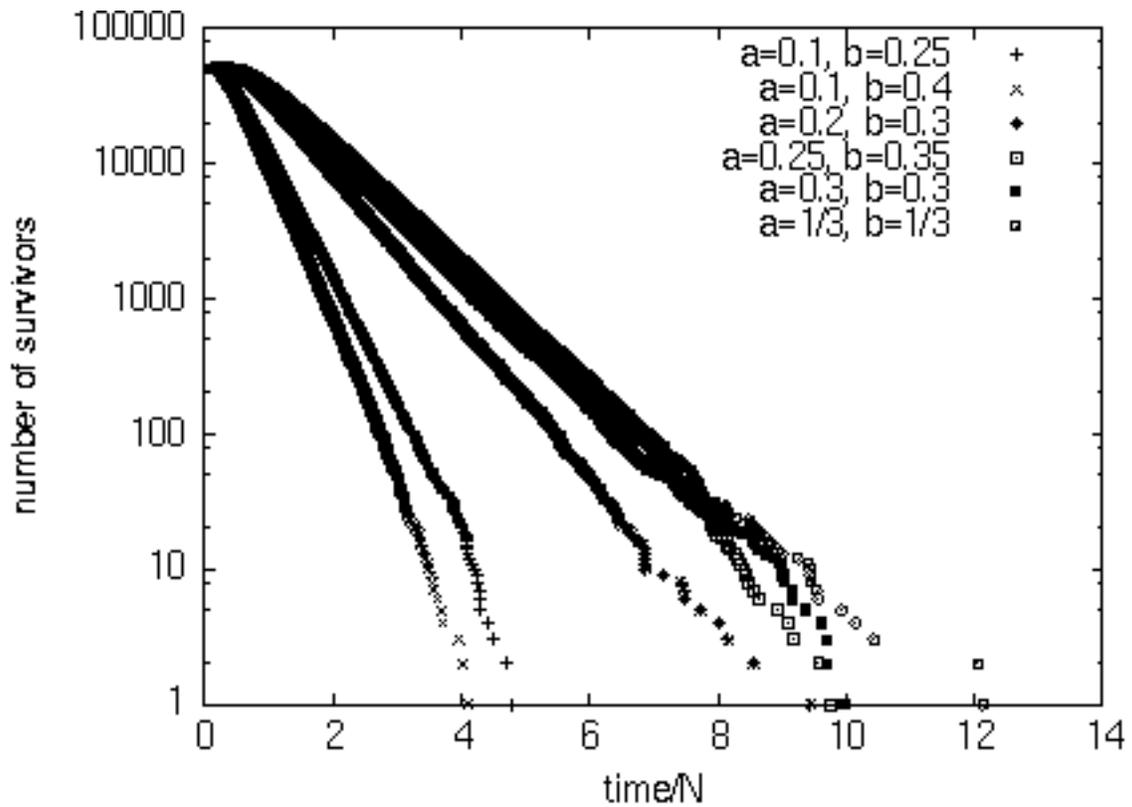}}
\caption{Number of survivors vs. time plots for the unequal rates cyclic 
competition case for system size 3000 and different combinations of 
$\alpha$ and $\beta$.}
\end{figure}

\end{document}